\documentclass[useAMS,usenatbib]{mnras}
\usepackage{amsfonts}
\usepackage{amsmath}
\usepackage{amssymb}
\usepackage{graphicx}
\usepackage{hyperref}
\usepackage{float}
\usepackage{color}

\newcommand{\Lsun}{\rm{\,L}_{\odot}}

\newcommand{\alf}{Alfv\'{e}n }

\newcommand{\imformat}{png}
\newcommand{\imformattwo}{pdf}

\newcommand{\highlight}[1]{#1} %Uncomment for no highlight version

\title[\textsc{shockfind}: MHD shocks in turbulent clouds]{SHOCKFIND - An algorithm to identify magnetohydrodynamic shock waves in turbulent clouds}
\author[Andrew Lehmann, Christoph Federrath and Mark Wardle]{Andrew Lehmann$^1$\thanks{E-mail:
andrew.lehmann@mq.edu.au}, Christoph Federrath$^2$ and Mark Wardle$^1$\\
$^1$Department of Physics and Astronomy, and Research Centre for Astronomy, Astrophysics \& Astrophotonics,\\
Macquarie University, Sydney, NSW 2109, Australia\\
$^2$Research School of Astronomy and Astrophysics,\\
Australian National University, Canberra, ACT 2611, Australia}

\begin{document}

%\pagerange{\pageref{firstpage}--\pageref{lastpage}} \pubyear{2002}
\pagerange{100--300} \pubyear{2002}

\maketitle

\begin{abstract}
The formation of stars occurs in the dense molecular cloud phase of the interstellar medium. \highlight{Observations and numerical simulations of molecular clouds have shown that supersonic magnetised turbulence plays a key role for the formation of stars}. Simulations have also shown that a large fraction of the turbulent energy dissipates in shock waves. The three families of MHD shocks --- fast, intermediate and slow --- distinctly compress and heat up the molecular gas, and so provide an important probe of the physical conditions within a turbulent cloud. \highlight{Here we introduce the publicly available algorithm, \textsc{shockfind}, to extract and characterise the mixture of shock families in MHD turbulence}. The algorithm is applied to a 3-dimensional simulation of a magnetised turbulent molecular cloud, and we find that both fast and slow MHD shocks are present in the simulation. We give the first prediction of the mixture of turbulence-driven MHD shock families in this molecular cloud, and present their distinct distributions of sonic and Alfv\'{e}nic Mach numbers.  Using subgrid one-dimensional models of MHD shocks we estimate that $\sim 0.03$~\% of the  volume of a typical molecular cloud in the Milky Way will be shock heated above 50~K, at any time during the lifetime of the cloud. We discuss the impact of this shock heating on the dynamical evolution of molecular clouds.
\end{abstract}


\section{Introduction} \label{ch:Intro}

The formation of stars is sensitive to the underlying physics of the supersonic magnetohydrodynamic (MHD) turbulence in molecular clouds \citep{mac_low_control_2004, elmegreen_interstellar_2004, krumholz_general_2005, mckee_theory_2007, hennebelle_analytical_2008, padoan_star_2011, hennebelle_turbulent_2012}. High resolution three-dimensional simulations of molecular clouds have shown that the star formation rate and efficiency depend on whether the turbulence is solenoidally or compressively driven \citep{federrath_star_2012,federrath_star_2013}, and that the stellar initial mass function is sensitive both to non-ideal MHD effects such as ambipolar diffusion \citep{mckee_sub-alfvenic_2010} and to the driving and Mach number of the turbulence \citep{hennebelle_analytical_2009, hennebelle_analytical_2013, hopkins_general_2013}. The importance of stellar feedback \citep{krumholz_radiation-hydrodynamic_2007, cunningham_radiation-hydrodynamic_2011, myers_star_2014, nakamura_protostellar_2007, wang_outflow_2010, price_effect_2008, price_inefficient_2009, offner_investigations_2014, federrath_modeling_2014, federrath_inefficient_2015, padoan_supernova_2015}, whether gravity drives turbulent motions \citep{elmegreen_accretion-driven_2010, klessen_accretion-driven_2010, vazquez-semadeni_molecular_2010, federrath_new_2011, robertson_adiabatic_2012} and the role that turbulence plays in producing the ubiquitously observed filaments \citep{arzoumanian_characterizing_2011, andre_filamentary_2014, smith_nature_2014, smith_nature_2016, federrath_universality_2016, kainulainen_gravitational_2016, hacar_musca_2016} are big questions that continue to be studied. Rigorous observational effects distinctly revealing the presence or dominance of the various physical processes are strongly sought after.

Observed non-thermal linewidths of molecular lines reveal a turbulence in molecular clouds that is highly supersonic, with Mach numbers typically in the range of 3 to 30 \citep{larson_turbulence_1981, solomon_mass_1987, ossenkopf_turbulent_2002, heyer_universality_2004, roman-duval_turbulence_2011, schneider_what_2013, henshaw_molecular_2016}. The supersonic flows in these clouds will inevitably form shock waves. MHD simulations by \cite{stone_dissipation_1998} found that $\sim 50 \%$ of the turbulent energy is dissipated to shocks. When turbulence is allowed to decay, a large range of weak shocks is responsible for the majority of the dissipation \citep{smith_shock_2000}, whereas a small range of stronger shocks dissipates the turbulence while it is continuously driven \citep{smith_distribution_2000}. If other turbulence parameters --- such as magnetic field strength, driving mechanism, Mach number, inclusion of other physical effects like self-gravity, stellar feedback, cosmic rays and others --- could be linked to distributions of shock waves, then the radiative signatures of shocks become observational diagnostics for these parameters in molecular clouds. The goal of this work is to determine this link, from turbulence parameters to distributions of shock waves, in \textit{simulations} of turbulent clouds in order to provide observational tests of the various physical processes.

MHD fluids can support three families of shocks: fast, intermediate and slow \citep{de_hoffmann_magneto-hydrodynamic_1950, kennel_mhd_1989}. Note that these names refer to the associated MHD linear wave modes, and not to the speed of a given shock. In Section~\ref{sec:shocks} we use the MHD jump conditions to detail the fundamental differences between the shock families. \cite{pon_molecular_2012} argue that, for molecular clouds, the turbulent cascade leads to low-velocity shocks (a few km/s) doing the majority of the dissipation. They show that even at low velocities, fast MHD shocks will radiate more strongly than photodissociation regions in mid-$J$ rotational transitions of CO. \cite{lehmann_signatures_2016} use two-fluid MHD models of shocks in molecular cloud conditions to show that low-velocity fast and slow MHD shocks distinctly compress and heat the molecular gas. At velocities less than 4~km/s slow shocks can reach compression ratios of up to 500 whereas fast shocks compress the gas less than 10 times the preshock values. In addition, slow shocks can reach peak temperatures up to $\sim 800$ K whereas fast shocks reach up to $\sim 150$ K. This is because in a weakly ionized gas --- such as in molecular clouds --- the ion-neutral collision timescale, which determines the heating timescale in fast shocks, is slower than the cooling timescale, which in turn is slower than the neutral-ion collision timescale that controls the heating in slow shocks. These higher peak temperatures in slow shocks result in stronger CO rotational lines (above $J = 6-5$) and low-lying pure rotational lines of H$_2$ ($\nu = 0-0$) than in fast shocks of the same velocity.

Molecular clouds are usually assumed to be in chemical equilibrium for the average conditions of the cloud. However, the inhomogeneous structure introduced by turbulence allows for local reaction rates to significantly differ from global average conditions \citep{hollenbach_molecular_1971, wolfire_neutral_1995, glover_modelling_2010}. Shocks are an important local mechanism for driving chemical reactions. \cite{kumar_astrochemical_2013} followed the chemical evolution of a parcel of gas through a turbulence simulation and found the abundances of molecules, like CH$_2$ and HCO, to be highly sensitive to shocked regions. Their work considers only hydrodynamic shocks and so cannot capture the qualitatively distinct behaviours of differing MHD shock families. For example, the higher temperatures of low-velocity slow shocks produce vastly different chemical abundances than fast shocks of the same velocities \citep{lehmann_signatures_2016}. Using a simple oxygen chemical network \citeauthor{lehmann_signatures_2016} show that while fast shocks leave preshock abundances mostly untouched, slow shocks can increase the abundances of molecules like OH, O$_2$ and H$_2$O by several orders of magnitude within the hot shock front. To understand the chemical makeup of turbulent molecular clouds it is therefore important to understand which families of MHD shocks are present and how much volume they affect.

While there is some observational evidence for the dissipation of molecular cloud turbulence in fast MHD shocks \citep{lesaffre_low-velocity_2013, pon_mid-j_2014, pon_mid-j_2016, larson_evidence_2015} and in slow MHD shocks \citep{lehmann_signatures_2016}, thus far no one has determined which kinds of MHD shocks are present in simulations of magnetised turbulent molecular clouds. In Section~\ref{sec:algorithm} we present an algorithm, \textsc{shockfind}, to detect and characterise the mixture of MHD shock types in such simulations. We apply our new algorithm to an MHD simulation of molecular cloud turbulence in Section~\ref{sec:turbulence}. Finally, we discuss these results and conclude our study in Sections~\ref{sec:discussion} and \ref{sec:conclusion} respectively. The \textsc{shockfind} algorithm, written in \textsc{python}, is publicly available and can be found on BitBucket (https://bitbucket.org/shockfind/shockfind) and the \textsc{python} Package Index (https://pypi.python.org/pypi/shockfind). It is released under the Apache license version 2.0, and comes with documentation including a user's guide.

\section{Magnetohydrodynamic Shocks}\label{sec:shocks}

\begin{figure}
  \centering
  	\includegraphics[width=0.51\textwidth]{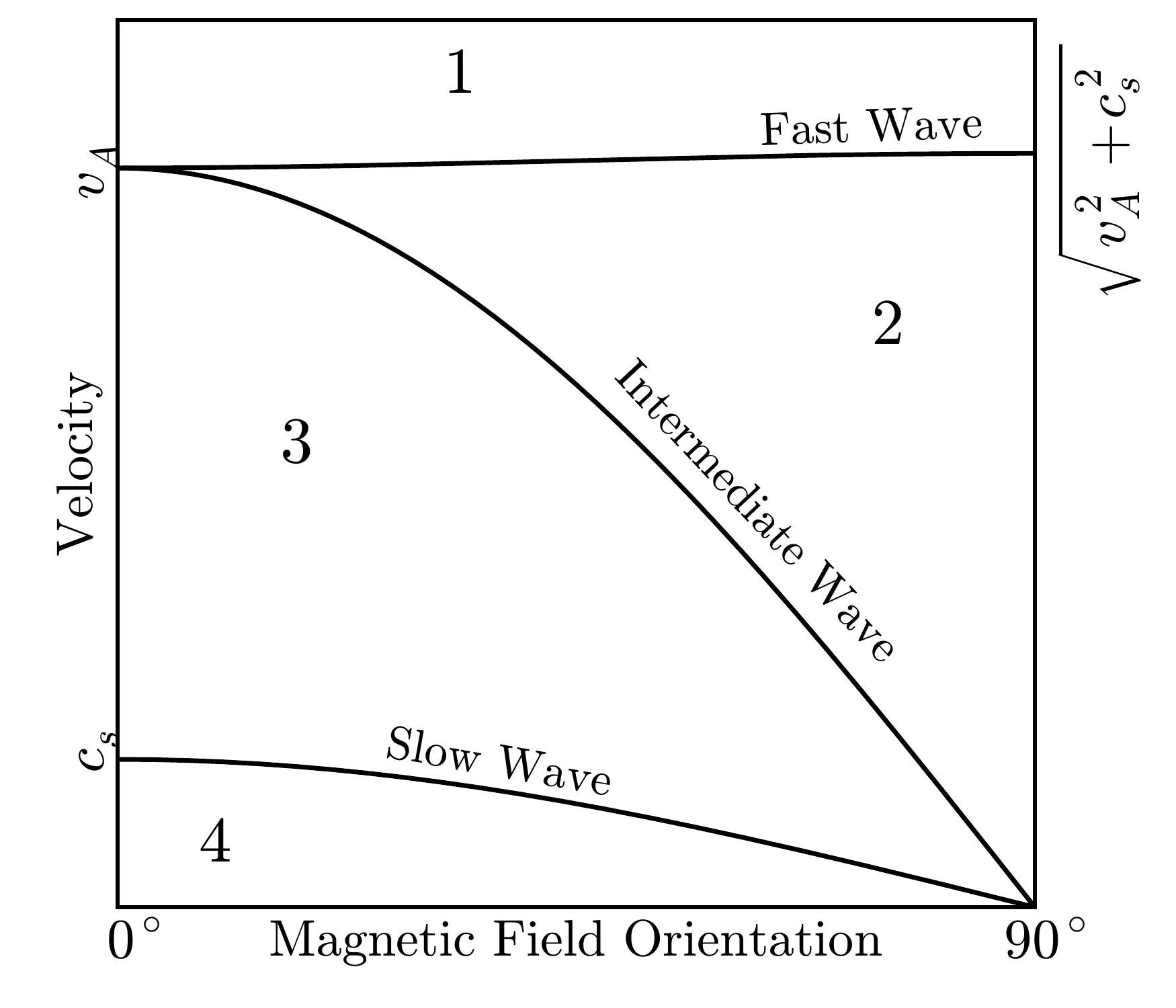}
 	\caption{The phase velocities of linear MHD wave modes versus the angle between the magnetic field and direction of propagation of those modes for $v_\mathrm{A} > c_\mathrm{s}$. The wave speeds delineate the regions marked $1$ to $4$.}
  \label{fig:ShockMap}
\end{figure}

Here we derive some fundamental differences between the shock families in the ideal limit of MHD. We summarise the relevant theoretical implications of the MHD jump conditions that can be found in \cite{lehmann_signatures_2016}, and illustrate how the various families of MHD shocks characteristically affect the ambient magnetic field.

In the frame of reference comoving with a shock wave --- the shock frame --- the pre-shock fluid has a speed, $v_\mathrm{s}$, greater than a linear wave speed in the fluid. The fluid then transitions inside a discontinuity to a fluid velocity less than a wave speed in the post shock fluid. In ideal MHD, the three linear waves supported, the fast, intermediate and slow waves have phase velocities
\begin{align*}
f &= \left( \frac{v_\mathrm{A}^2 + c_\mathrm{s}^2}{2} + \frac{1}{2}\sqrt{\left( v_\mathrm{A}^2 + c_\mathrm{s}^2 \right)^2 - 4v_\mathrm{A}^2c_\mathrm{s}^2\cos ^2\theta} \right)^{1/2} ,\\
i &= v_\mathrm{A} \cos \theta ,\\
s &= \left( \frac{v_\mathrm{A}^2 + c_\mathrm{s}^2}{2} - \frac{1}{2}\sqrt{\left( v_\mathrm{A}^2 + c_\mathrm{s}^2 \right)^2 - 4v_\mathrm{A}^2c_\mathrm{s}^2\cos ^2\theta} \right)^{1/2},
\end{align*}
where $v_\mathrm{A}=B/\sqrt{4 \pi \rho}$ is the \alf velocity, $c_s = \sqrt{k_\mathrm{B} T/ \mu_\mathrm{m}}$ is the isothermal sound speed with Boltzmann constant $k_\mathrm{B}$ and mean mass per particle $\mu _\mathrm{m}$, and $\theta$ is the angle between the magnetic field and the direction of propagation of the wave. These speeds are plotted as functions of $\theta$ in Fig.~\ref{fig:ShockMap} for $v_\mathrm{A} > c_\mathrm{s}$, as is usually the case in molecular clouds.

The three wave speeds demarcate four regions of fluid velocities marked 1 to 4 in Fig.~\ref{fig:ShockMap}. There are six ways to transition across at least one wave speed within the shock front, resulting in three families of MHD shocks: fast, intermediate and slow shocks. Fast shocks cross the fast wave speed only (1--2), intermediate shocks cross the intermediate wave speed (1--3, 1--4, 2--3, and 2--4), and slow shocks cross the slow wave speed only (3--4). This means that for a fast shock the Alfv\'{e}nic Mach number --- defined by $\mathcal{M}_\mathrm{A} \equiv v_\mathrm{s} / v_\mathrm{A}$ --- is necessarily greater than unity, whereas for a slow shock the shock speed is sub-Alfv\'{e}nic, i.e., $M_A < 1$. We will use this criterion to distinguish fast from slow shocks in the next section.

\begin{figure}
  \centering
    \includegraphics[width=0.48\textwidth]{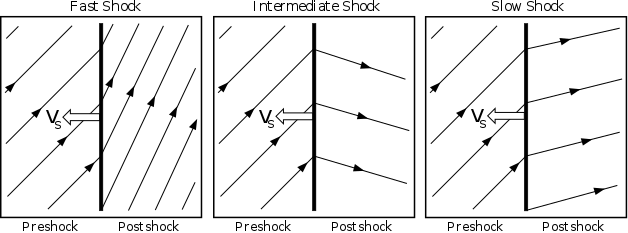}
    \caption{The effect on magnetic field orientation of the three classes of MHD shock waves. Fast shocks (left) increase the angle between the field and shock normal, intermediate shocks (middle) reverse the sign of the angle and slow shocks (right) decrease it. Hence the magnetic field strengthens across fast shocks and weakens across slow shocks.}
  \label{fig:ShockClasses}
\end{figure}

For a time-independent, plane-parallel shock wave, the pre- and post-shock mass density, velocity, gas pressure and magnetic field are related by the Rankine-Hugoniot jump conditions \citep{kennel_mhd_1989}. A consequence of these jump conditions are changes in magnetic field geometry across a shock front characteristic of each shock family. In fast shocks, the component of the magnetic field perpendicular to the direction of propagation, $B_\perp$, must increase from the pre-shock to the post-shock value. In intermediate shocks, $B_\perp$ must switch sign. This switch is due to a rotation of the magnetic field within the shock front. Finally, in slow shocks $B_\perp$ must decrease. For all shocks, the planar symmetry implies that the magnetic field parallel to the direction of propagation is conserved across the shock front. These three characteristic changes of the magnetic field direction are shown schematically in Fig.~\ref{fig:ShockClasses}.

The magnetic field strength is proportional to the separation of field lines, so one can see from Fig.~\ref{fig:ShockClasses} that the field strength increases across fast shocks and decreases across slow shocks. We will use this fact as a signature of these two classes of shocks in Section~\ref{sec:algorithm}. This also means that some of the kinetic energy of a fast shock is converted into magnetic field energy. Hence, for slow shocks at the same velocity as fast shocks, a greater portion of the energy budget is available to heat the gas. \highlight{As we do not account for the magnetic field geometry, but rather use only the field strength as an indicator of shock type we do not explicitly capture switch-on or switch-off shocks. However, these shocks are special cases of fast and slow shocks and so are counted amongst these categories. Finally, the magnetic field strength in intermediate shocks can either increase or decrease across the shock front depending on the initial conditions of a particular shock. Hence there is no simple signature of intermediate shocks in the magnetic field strength. In addition, there has been debate over whether intermediate shocks are physically admissable \cite[e.g.,][]{wu_mhd_1987, falle_inadmissibility_2001}, and so this class of MHD shocks is not considered further in this work.}

The structure of the magnetic field across the different shock waves only depends on the ideal MHD jump conditions. The heating inside the shock front, however, can depend on non-ideal effects. For example, in molecular clouds the gas is weakly ionized and so ion and neutral species can be decoupled. In this case, a multi-fluid approach is necessary to model the structure of the shock front. In fast shocks, the strong magnetic pressure behind the shock front drives ion species ahead of the neutrals in a magnetic precursor \citep{mullan_structure_1971, draine_interstellar_1980}. Collisional heating in fast shocks is thus controlled by the ion-neutral collision timescale, which is generally larger than the cooling timescale for low-velocity shocks (a few km/s) in molecular clouds. If the cooling keeps the temperature low within the shock, the neutral velocity may remain supersonic throughout and the fluid variables will smoothly transition from pre-shock to post-shock values in what is called a C-type shock. In fast shocks with shock velocity exceeding the magnetosonic speed of the charged fluid, defined by
\begin{align*}
v_m = \sqrt{c_\mathrm{s}^2 + v_{A,c}^2}
\end{align*}
where $v_{A,c}$ is the Alfv\'{e}n velocity in the charged fluid, a thin jump will form in which the heating is determined by molecular viscosity. Such a fast shock is called J-type. We see then, that we need to determine the local pre-shock conditions in turbulent molecular clouds in order to predict the shock heating from fast MHD shocks.

In the two-fluid models of \cite{lehmann_signatures_2016} slow shocks produce a temperature structure distinct from fast shocks. Shocks of this family are driven by the gas pressure of the neutrals, and so heating is determined by the neutral-ion collision timescale. This timescale is shorter than the cooling timecale and so high peak temperatures are reached in a thin shock front resembling an ordinary hydrodynamic shock. The jump in temperature across this shock front is determined by the sonic Mach number $\mathcal{M} \equiv v_\mathrm{s}/c_\mathrm{s}$:
\begin{align}
\frac{T_2}{T_1} &= \left( 1 + \frac{2 \gamma}{\gamma + 1}\left( \mathcal{M}^2 -1 \right) \right) \frac{\mathcal{M}^2 \left( \gamma -1 \right) +2}{\mathcal{M}^2\left( \gamma +1 \right)}\label{eq:slowtemp} 
\end{align}
which reaches 200 K at low shock velocities ($v_\mathrm{s}\sim$2~km/s for $c_\mathrm{s}=0.2$~km/s) and adiabatic index $\gamma = 5/3$. This is hot enough to produce chemical abundances and molecular cooling significantly different to ambient molecular cloud conditions.

Fast and slow shocks therefore present distinct temperature structures within molecular clouds. To fully understand the heating and chemistry driven by turbulent dissipation in shock waves it is therefore critical to determine the relative fraction of shock families. We will do this in the following by using numerical MHD simulations of turbulent molecular clouds.

\section{Shock Detection Algorithm}\label{sec:algorithm}

In this section, we present a new algorithm, \textsc{shockfind}, to detect and characterise the shocks in MHD simulations of molecular clouds. We then test this algorithm by considering a case study simulation of colliding MHD shocks, which should be a common occurence in supersonic MHD turbulence.

\subsection{Algorithm Summary}\label{sec:alg_summary}

Here we summarise the seven-step algorithm, \textsc{shockfind}, for detecting fast or slow shocks in an MHD simulation. For a simulation with mass density $\rho$, three velocity components $u_x$, $u_y$, and $u_z$, and three magnetic field components $B_x$, $B_y$, and $B_z$, the algorithm will:

\begin{enumerate}

\item Identify shock candidates as computational cells with large convergence:
\begin{align}
-\nabla \cdot \mathbf{u} = - \left( \partial_x u_x + \partial_y u_y + \partial_z u_z \right) \label{eq:convergence}
\end{align}
or large magnitude of the density gradient:
\begin{align}
|\nabla \rho| = \sqrt{\left( \partial_x \rho \right)^2 + \left( \partial_y \rho \right)^2 + \left( \partial_z \rho \right)^2 }
\end{align}
where large is above a user-defined threshold appropriate to the particular simulation;

\item Compute the shock direction at the location of each candidate cell, $\mathbf{n}_\mathrm{s}$, using the gradient of the density:
\begin{align}
\mathbf{n}_\mathrm{s} = \nabla \rho / \left| \nabla \rho \right| = \left( \partial_x \rho, \, \partial_y \rho, \, \partial_z \rho\right) / \left| \nabla \rho \right|;
\end{align}

\item Extract averaged fluid variables along a cylinder perpendicular to the shock front. The cells, at coordinates $\mathbf{r}$, on the central axis of the cylinder are defined by
\begin{align}
\mathbf{r} = \mathbf{r}_\mathrm{s} + \lambda \mathbf{n}_\mathrm{s}
\end{align}
where $\mathbf{r}_\mathrm{s}$ is the location of the candidate cell and $\lambda$ parametrises the line and ranges from $\pm$ a few shock thicknesses, $N$, that the simulation spreads the shock over. The effect of varying the radius of this cylinder is shown in Appendix~\ref{ap:cylinder};

\item The cell-averaged variables for $\lambda > N/2$ are the pre-shock values, while the cell-averaged variables for $\lambda < -N/2$ represent the post-shock values;

\item Compute the shock speed using the pre- and post-shock parallel velocities and densities obtained in step (iv): 
\begin{align}
v_\mathrm{s} &= \left( u_{\parallel,\mathrm{pre}} - u_{\parallel,\mathrm{post}} \right) / \left( 1 - \rho_\mathrm{pre}/\rho_\mathrm{post} \right)\label{eq:shockspeed}
\end{align}
where $u_\parallel=\mathbf{u} \cdot \mathbf{n}_\mathrm{s}$;

\item Compare the shock speed $v_\mathrm{s}$ to the pre-shock Alfv\'{e}n velocity to form the Alfv\'{e}nic Mach number $\mathcal{M}_\mathrm{A}$. Fast shocks must have $\mathcal{M}_\mathrm{A} > 1$ and slow shocks must have $\mathcal{M}_\mathrm{A} < 1$. In addition, we compare the pre- and post-shock magnetic field strengths. Fast shocks must have $B_\mathrm{post}/B_\mathrm{pre}>1$ and slow shocks must have $B_\mathrm{post}/B_\mathrm{pre}<1$. If a shock candidate consistently satisfies both of these inequalities then we have detected a fast or slow MHD shock, otherwise it is only a candidate and is not included in the further analysis;

\item Finally, we filter the detected shock cells by ignoring those detections that do not occur at local maxima in the convergence along the extracted line of step (iii). This step avoids extracting multiple cells as individual shocks that actually belong to the same shock. This process is illustrated in Appendix~\ref{ap:overcounting}.
\end{enumerate}

\subsection{Test simulation of colliding MHD shocks}

\begin{figure}
  \centering
    \includegraphics[width=0.5\textwidth]{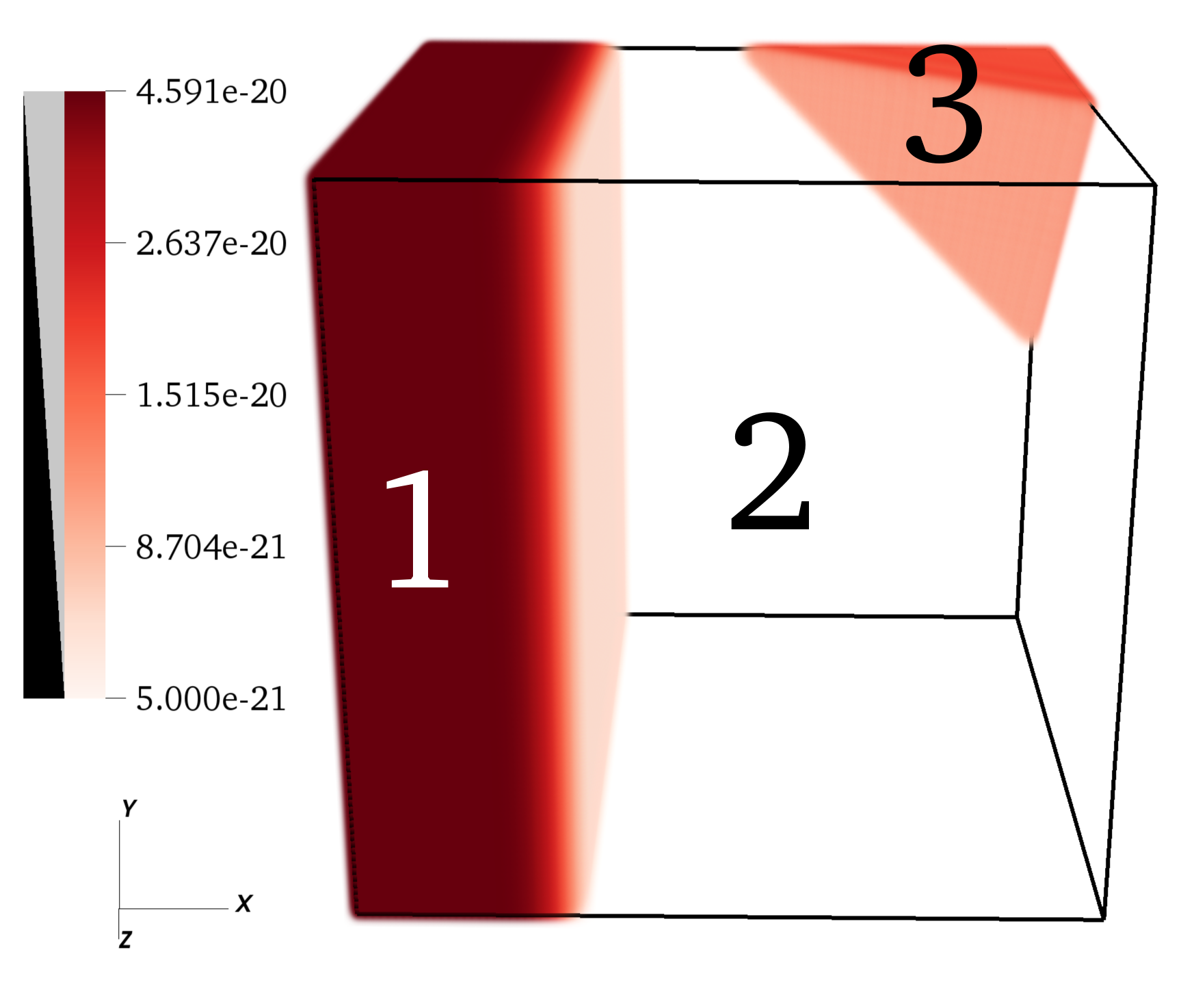}
    \caption{Initial density configuration of the colliding shock simulation. The regions labeled are (1) the slow post-shock region, (2) the common pre-shock region and (3) the fast post-shock region.}
  \label{fig:smash3d}
\end{figure}

\begin{table*}
{\centering
\caption{Initial setup of colliding MHD shock simulation}
\begin{tabular}{l c c c} \hline
\multicolumn{4}{c}{\vspace{-0.4cm}}\\
Variable & Slow Post-shock (1) & Common Pre-shock (2) & Fast Post-shock (3)\\
\multicolumn{4}{c}{\vspace{-0.4cm}}\\ \hline
$\rho \, (10^{-20}\,\mathrm{g/cm}^3)$ & $4.591$ & $0.385$ & $1.800$\\
$p \, (10^{-11}\,\mathrm{g\,cm / s}^2)$ & $5.995$ & $0.154$ & $11.48$\\
$u_x \, (\mathrm{km/s})$ & $3.803$ & $2.887$ & $1.141$\\
$u_y \, (\mathrm{km/s})$ & $3.868$ & $2.887$ & $1.141$\\
$u_z \, (\mathrm{km/s})$ & $-2.887$ & $-2.887$ & $0.430$\\
$B_x \, (\mu\mathrm{G})$ & $24.75$ & $24.75$ & $69.07$\\
$B_y \, (\mu\mathrm{G})$ & $5.569$ & $24.75$ & $69.07$\\
$B_z \, (\mu\mathrm{G})$ & $0$ & $0$ & $88.64$\\ \hline
\end{tabular}
\label{tab:fast-slow}
}
\end{table*}

\begin{figure*}
  \centering
    \includegraphics[width=1.05\textwidth]{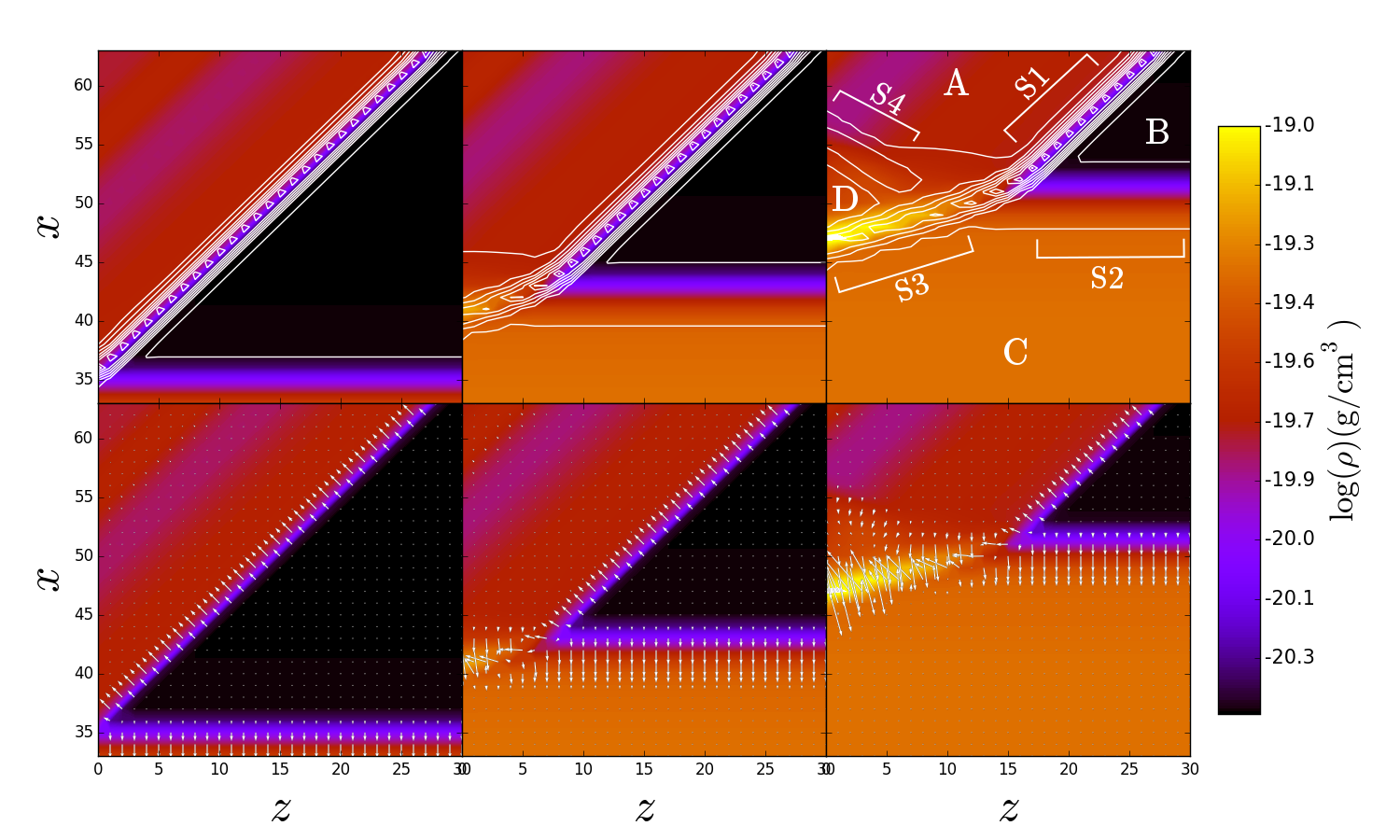}
    \caption{Three snapshots of mass density slices (constant $y$) from the colliding MHD shocks simulation. The colours show mass density. In the upper row, the white contours are convergence ($-\nabla \cdot \mathbf{v}$) in this plane. In the lower row, vectors are projected gradient of density ($\nabla \rho$). The labels A--D denote different density regions and S1--S4 are shock candidates detailed in the text.}
  \label{fig:smash-sequence}
\end{figure*}

In this section we illustrate the steps of our shock detection algorithm outlined in Section 3.1, by applying it to a simulation of colliding shock waves. We use the code FLASH \citep{fryxell_flash_2000, dubey_challenges_2008} in version 4 to integrate the ideal, three-dimensional MHD equations. The equations are solved on a grid with a total of $128^3$ grid cells using the HLL3R positive-definite Riemann solver \citep{waagan_robust_2011}. The equations are closed with an ideal gas equation of state with adiabatic index $\gamma = 1.1$.

In this simulation we initialise a box with a slow MHD shock with shock speed $v_\mathrm{s} = 1$~km/s travelling in the direction $n_\mathrm{s} = (1,0,0)$, and a fast MHD shock with shock speed $v_\mathrm{s} = 5$~km/s travelling in the direction $n_\mathrm{f} = (-1,-1,1)$. We choose pre-shock variables (common to both shocks) of density $\rho = 3.85 \times 10^{-21} \, \mathrm{g / cm}^3$, pressure $p = 1.54 \times 10^{-12} \, \mathrm{g \, cm/s}^2$, and a magnetic field strength of 35~$\mu$G oriented at $45^\circ$ to the slow shock front. These pre-shock values are used to compute post-shock values using the MHD jump conditions \citep{kennel_mhd_1989}, and we choose a frame of reference such that the fast shock is stationary. Fig.~\ref{fig:smash3d} shows the initial density configuration of the three regions --- slow post-shock, common pre-shock, and fast post-shock --- and the fluid variables are listed in Table~\ref{tab:fast-slow}.

\subsubsection{Convergence and density gradient}

In Fig.~\ref{fig:smash-sequence} we plot slices of mass density of the colliding shock simulation at three times in the simulation. The rightmost column shows a slice after the two shocks have collided and a significant interaction region has developed. The coloured contours show the density with 4 regions marked A--D. Region A is the post-shock region of the initial 5 km/s fast shock, region B is the common pre-shock region, region C is the post-shock region of the initial 1 km/s slow shock, and finally region D is the post-interaction region.

Overplotting the density contours in the upper row of Fig.~\ref{fig:smash-sequence} are white line contours of strong convergence, defined by equation~\eqref{eq:convergence}. This quantity distinctly picks out candidate shock fronts labelled S1-S4. S1 and S2 are the initial fast and slow shocks we set up to collide. S3 and S4 are the results of the collision, which have geometries suggestive of a refractive (S3) and reflective (S4) process.

In the lower row of Fig.~\ref{fig:smash-sequence} we plot projected vectors of the gradient of the density over the contours of density. This vector points in the direction of increasing density and so it always points towards the plane of a shock front. This allows us to define a line through a shock at which we extract the fluid variables. It also allows us to compute the fluid velocity in the direction of shock propagation, $u_\parallel$, by projecting onto the direction given by the gradient.

\subsubsection{Shock family criteria}

\begin{figure}
  \centering
    \includegraphics[width=0.47\textwidth]{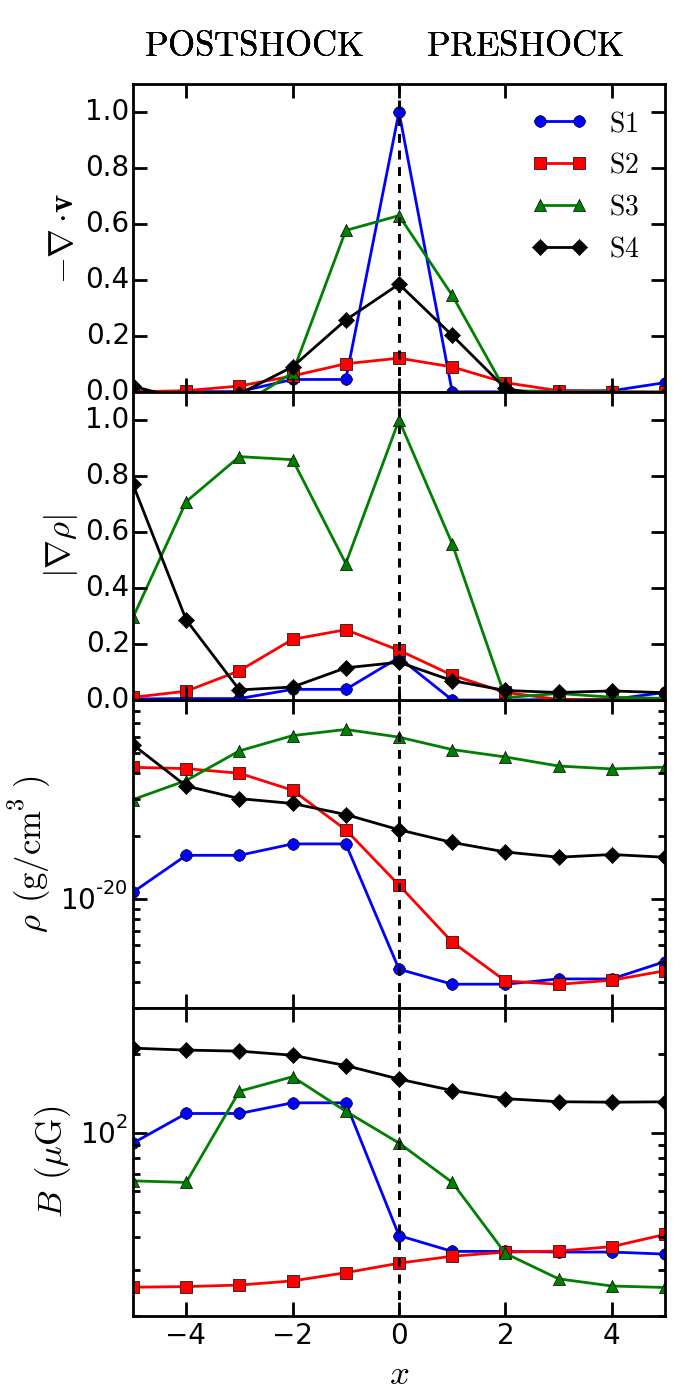}
    \caption{Line profiles through the shock candidates S1--S4 from Fig.~\ref{fig:smash-sequence}. The profiles are of convergence ($-\nabla \cdot \mathbf{v}$), magnitude of the gradient of the mass density, mass density, and magnetic field strength. The convergence and gradient are normalised to their peak values. The x axis is in units of grid cell lengths.}
  \label{fig:fourplot}
\end{figure}

Using the convergence and gradient as described above, we plot in Fig.~\ref{fig:fourplot} extracted fluid variables through the shock candidates S1-S4 from Fig.~\ref{fig:smash-sequence}. The $x$-axis is zeroed at the convergence peak, with the post-shock region at negative values of $x$ and pre-shock region at positive values of $x$. The convergence and gradient are normalised to their peak values. To compare pre- and post-shock fluid variables we take the average of the variable over a few cells either side of the convergence peak.

The S1 and S2 shocks (which were the initial fast and slow shocks, respectively) show strong density contrasts though the simulation has spread out the S2 shock (red lines in Fig.~\ref{fig:fourplot}) over a wider range than the S1 shock. This is also reflected in a much lower and wider convergence peak in the slow shock. As discussed in Section~\ref{sec:shocks} the magnetic field, the lower panel of Fig.~\ref{fig:fourplot}, in the S1 shock is stronger in its post-shock region compared to its pre-shock region because it is a fast MHD shock. Conversely, in the S2 shock $B$ is stronger in the pre-shock region because it is a slow MHD shock.

Once we have computed pre- and post-shock densities, $\rho_1$ and $\rho_2$, respectively, and the pre- and post-shock parallel velocities, $u_1$ and $u_2$, respectively, we can determine the shock velocity $v_\mathrm{s}$ using equation~\eqref{eq:shockspeed}. By comparing the shock speed to the Alfv\'{e}n velocity we form the Alfv\'{e}nic Mach number $\mathcal{M}_\mathrm{A}$. As discussed in Section~\ref{sec:shocks}, for fast shocks $\mathcal{M}_\mathrm{A} > 1$ and for slow shocks $\mathcal{M}_\mathrm{A} < 1$. For the S1 and S2 candidates, we find that $\mathcal{M}_\mathrm{A} \sim 3.1$ and $\sim 0.6$ respectively, confirming their status as fast and slow MHD shocks.

The S3 and S4 shock candidates both have magnetic field stronger in the post-shock region than in the pre-shock region, indicating that they are fast MHD shocks. Indeed, for S3 the Alfv\'{e}nic Mach number $\mathcal{M}_\mathrm{A} \sim 16.5$ and so it satisfies both criteria. However, for the S4 candidate $\mathcal{M}_\mathrm{A} \sim 0.8$, ruling it out as a fast shock. As it is a planar converging structure it could be a fast linear wave or even an intermediate shock, but our simple criteria cannot distinguish these cases. We will call converging structures that show magnetic field ratios inconsistent with Alfv\'{e}nic Mach number criteria \textit{fast-like} and \textit{slow-like} disturbances and exclude them from further analyses.

\section{Molecular Cloud Turbulence}\label{sec:turbulence}

\begin{figure*}
  \centering
    \includegraphics[width=\textwidth]{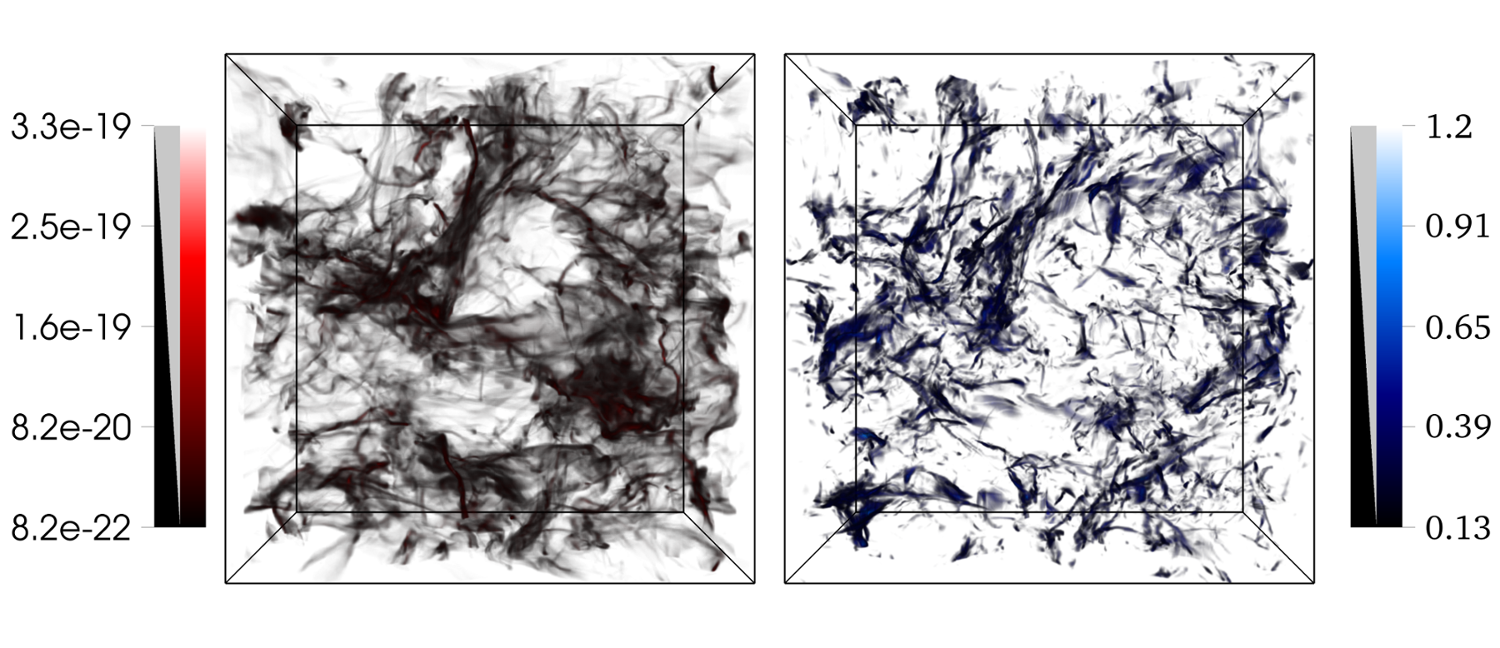}
    \caption{3D renderings of the MHD simulation of turbulence. (Left) mass density above the average mass density $\left<\rho\right> = 8.2\times 10^{-22} ~ \mathrm{g/cm}^3$. (Right) Convergence ($-\nabla \cdot \mathbf{v}$) normalised by the ratio of the velocity dispersion to the cell size.}
  \label{fig:turbulence}
\end{figure*}

Here we present an application of our shock-detection algorithm to a 3D simulation of molecular cloud turbulence. The ultimate goal is to determine the fraction of slow and fast shocks and their respective effects for the heating and evolution of molecular clouds.

We use the turbulent initial conditions of molecular cloud simulation model 21 (GT256mM10B10) from \cite{federrath_star_2012}. It is a 3D simulation of an isothermal, ideal MHD, turbulent molecular cloud with mixed compressive and solenoidal driving. \highlight{The turbulence is driven to maintain a velocity dispersion $\sigma \sim 1.8$ km/s. As the sound speed $c_s = 0.2$, the turbulence contains a large fraction of supersonic gas and is therefore expected to drive shocks with Mach numbers $\mathcal{M}\sim 9$}. The initial magnetic field strength is 10 $\mu$G, and the time step that we analyse here occurs after the turbulence has been fully developed but before self-gravity has been turned on to study star formation. The details of the integration scheme can be found in that paper.

Fig.~\ref{fig:turbulence} shows three-dimensional renderings of the mass density ($\rho$, left panel) and convergence ($-\nabla \cdot \mathbf{v}$, right panel) of the simulation cube. The mass density is cut off at the average value $\left< \rho \right> = 8.2\times 10^{-22} ~ \mathrm{g/cm}^3$, so that we plot only the high density regions. These regions are highly filamentary and fill only a small volume of the cloud. The convergence has been normalised by $\sigma/\Delta x$ where $\sigma$ is the 3D velocity dispersion of the cloud and $\Delta x$ is the cell size. There is a rough correlation between regions of strong convergence and regions of high density. This would be expected if the highest densities in turbulent clouds are post-shock layers.

\subsection{Search Thresholds}\label{sec:thresholds}

\begin{figure}
  \centering
    \includegraphics[width=0.48\textwidth]{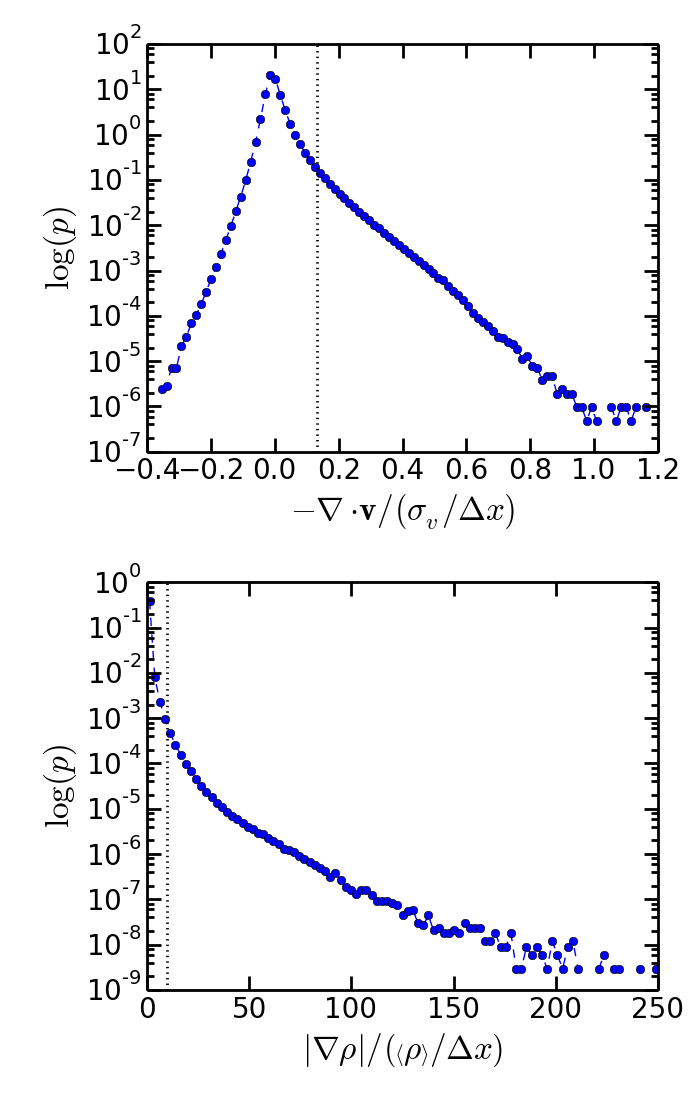}
    \caption{Distribution of (upper) normalised convergence and (lower) normalised magnitude of gradient. The dashed vertical lines are the search thresholds, which estimates the convergence and density gradient of a shock with velocity $v_\mathrm{s} = 1$ km/s, \highlight{pre-shock density $\rho_0 = 10\left< \rho \right>$ and compression ratio $r=4$ spread over 3 cells (see text for details of estimate).}}
  \label{fig:thresholds}
\end{figure}

While \textsc{shockfind} could check for shocks at every cell in the simulation, it would be computationally expensive to check all $512^3$ cells of this simulation. Considering that non-converging cells can be ruled out as shock candidates without further analysis, we develop search criteria to speed up the process. 

Fig.~\ref{fig:thresholds} shows the probability distribution functions (PDFs) of the convergence (upper) and the magnitude of the gradient of density (lower). The magnitude of the gradient of density has been normalised by $\left<\rho \right>/\Delta x$ where $\left<\rho \right>$ is the average mass density. We can estimate the convergence and gradient across a shock wave with shock velocity $v_\mathrm{s}$ propagating into a gas with pre-shock mass density $\rho_0$ as
\begin{align*}
\left(-\nabla \cdot \mathbf{v}\right)_\mathrm{s} &\sim - \frac{v_2 - v_\mathrm{s}}{N\Delta x}\\
\left(\nabla \rho\right)_\mathrm{s} &\sim \frac{\rho_2 - \rho_0}{N\Delta x}
\end{align*}
where $v_2$ is the post-shock velocity, $\rho_2$ is the post-shock mass density and $N$ is the number of cells the simulation typically spreads a discontinuity over. We use N=3 in this work. Using the relation $\rho_2 v_2=\rho_0 v_\mathrm{s}$ and normalising as above, these estimates become
\begin{align*}
\frac{\left(-\nabla \cdot \mathbf{v}\right)_\mathrm{s}}{\sigma/\Delta x} &\sim \frac{v_\mathrm{s}}{\sigma} \frac{r - 1}{r N }\\
\frac{\left(\nabla \rho\right)_\mathrm{s}}{\left<\rho \right>/\Delta x} &\sim \frac{\rho_0 }{\left<\rho \right>}\frac{r - 1}{N}
\end{align*}
where $r=\rho_2/\rho_0$ is the compression ratio. \highlight{We use the compression ratio $r$ to control the search thresholds of step (i) of the algorithm (Section~\ref{sec:alg_summary}). In Fig.~\ref{fig:thresholds} the dashed vertical lines show the thresholds for shock velocity $v_s = 1$~km/s, compression ratio $r=4$, and pre-shock density $\rho_0 = 10 \left< \rho \right>$. As these thresholds are treated independently (Step~(i) of Section~\ref{sec:alg_summary}), cells that do not satisfy one threshold may still be identified as a shock candidate if they satisfy the other threhold. This conservative approach means we look at more cells than if we applied both thresholds simultaneously. At a velocity of 1 km/s, a slow shock will reach a peak temperature of 55 K (equation~\eqref{eq:slowtemp}). Molecular cooling is more efficient at high temperatures and high densities, and so running the algorithm at cells above these thresholds ensures we extract the most observationally relevant shocks.}

\subsection{Shock family statistics}

\begin{figure*}
  \centering
    \includegraphics[width=0.7\textwidth]{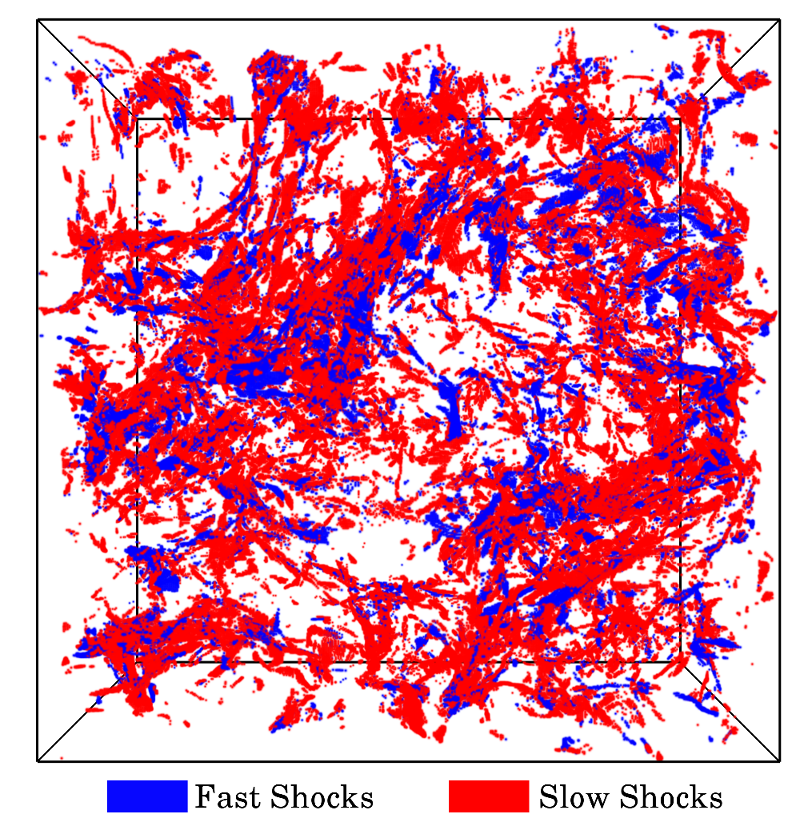}
    \caption{Spatial distribution of the shocks that we detected with our new shock-detection algorithm \textsc{shockfind}.
  \label{fig:shockresults}}
\end{figure*}

Using the thresholds defined in the previous section, the spatial distribution of fast and slow shocked cells is shown in Fig.~\ref{fig:shockresults}. Red points refer to slow shocks and blue points refer to fast shocks. Many of the detected cells form connected shock front sheets and others form long filamentary structures. Around 40\% of searched cells fail to consistently satisfy the \highlight{Alfv\'{e}nic} Mach number and magnetic field ratio criteria (Step~(vi) of Section~\ref{sec:alg_summary}). A further half of the detected shocked cells are filtered out because they do not lie on local maxima in convergence (Step~(vii) of Section~\ref{sec:alg_summary}). We analyse the results of this search in the following sections.

\begin{figure}
  \centering
    \includegraphics[width=0.5\textwidth]{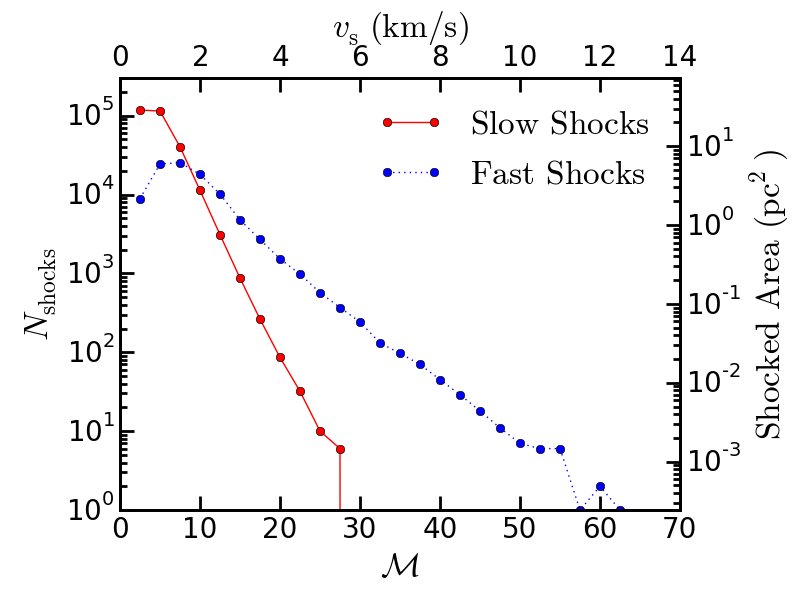}
    \caption{Distribution of sonic Mach numbers in bins centred on every $v_\mathrm{s}=0.5$ km/s with size 0.5 km/s. The blue (dotted) line refers to fast shocks and red (solid) line refers to slow shocks. The right axis shows the area that the shock fronts occupy.}
  \label{fig:histospeeds}
\end{figure}

\begin{figure}
  \centering
    \includegraphics[width=0.5\textwidth]{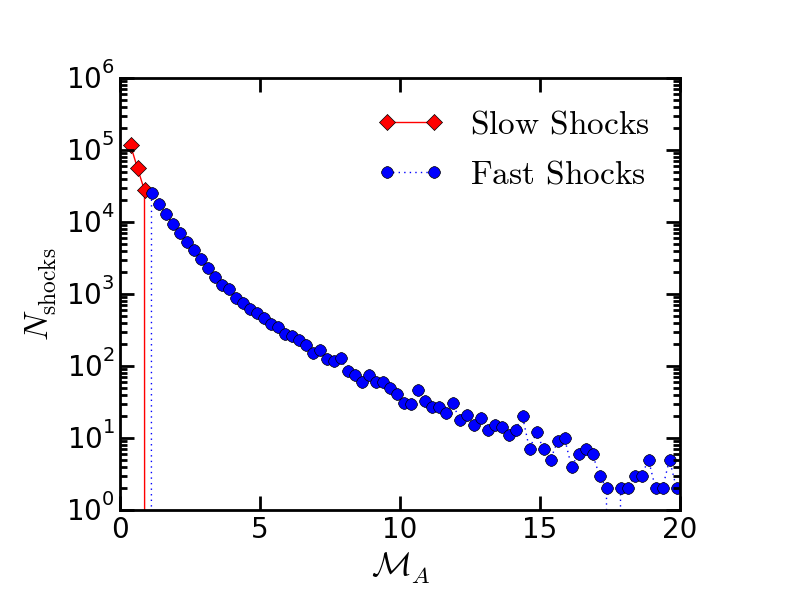}
    \caption{Distribution of Alfv\'{e}nic Mach numbers, with bin boundaries every 0.25. The blue (dotted) line refers to fast shocks and red (solid) line refers to slow shocks.}
  \label{fig:histomachalf}
\end{figure}

Fig.~\ref{fig:histospeeds} shows the distributions of sonic Mach numbers for fast shocks (blue) and slow shocks (red). The slow shock distribution steeply and monotonically decreases, with a larger number of slow shocks than fast below $\mathcal{M}\sim 8$ ($v_\mathrm{s} \sim 1.5$~km/s). The fast shock distribution peaks around $\mathcal{M}\sim 10$ before slowly decreasing. We estimate the area occupied by the shock fronts by treating each detected shocked cells as having an area of one of its faces: $\left(\Delta x \right)^2 \sim 2.4 \times 10^{-4}$~pc$^2$. \highlight{We perform a convergence test shown in Appendix~\ref{ap:convergence}. While there are more very low velocity shocks ($\mathcal{M} < 5$) to be found below our thresholds defined in Section~\ref{sec:thresholds}, the Mach number distributions are well converged for the most observationally relevant shocks.} Fig.~\ref{fig:histomachalf} shows the distributions of Alfv\'{e}nic Mach numbers for fast shocks (blue) and slow shocks (red). By definition slow shocks are sub-Alfv\'{e}nic and fast shocks are super-Alfv\'{e}nic and so the distributions distinctly lie on either side of unity. In the two-fluid MHD shocks of \cite{lehmann_signatures_2016} the peak temperature of slow shocks is determined by the sonic Mach number, whereas for fast shocks it is determined by the competition of molecular cooling and ion-neutral collisional heating. In Section~\ref{sec:heating} we use these two distributions (Figs.~\ref{fig:histospeeds} and \ref{fig:histomachalf}) to make an estimate of turbulence-driven shock heating.

\begin{figure}
  \centering
    \includegraphics[width=0.5\textwidth]{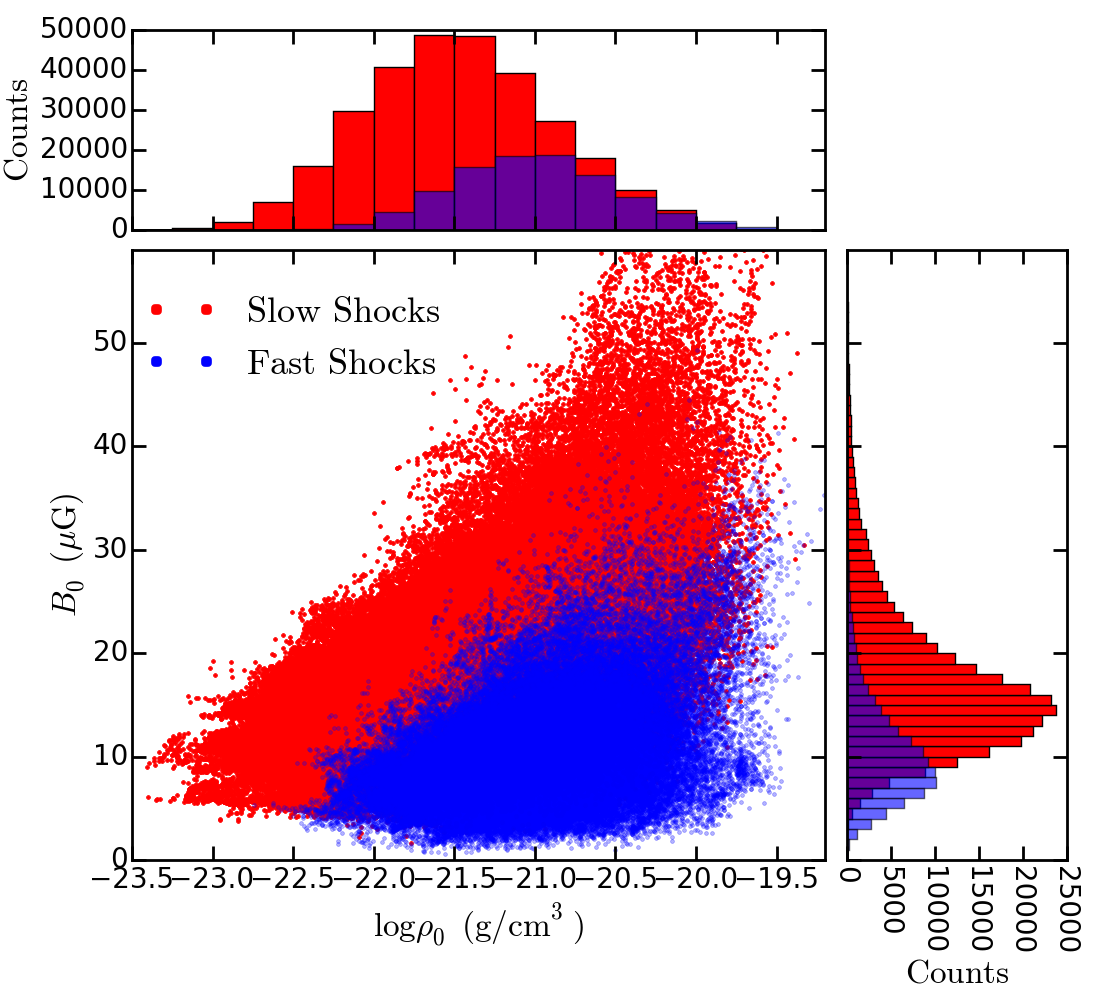}
    \caption{Distribution of pre-shock magnetic field strengths and mass densities.}
  \label{fig:BversusRHO}
\end{figure}

In Fig.~\ref{fig:BversusRHO} we plot the distributions of pre-shock magnetic field strengths and mass densities for the search described above. These distributions give us the typical pre-shock variables that should be used to model shocks relevent to molecular cloud turbulence. They show that the typical pre-shock conditions are different for fast and slow shocks. For example, for fast shocks the average pre-shock mass density $\left< \rho_0 \right>_\mathrm{f} \sim 2\times 10^{-21}$~g/cm$^3$ whereas for slow shocks $\left< \rho_0 \right>_\mathrm{s} \sim 9\times 10^{-22}$~g/cm$^3$. This corresponds to total hydrogen densities, $n_\mathrm{H} = n(\mathrm{HI}) + 2n(\mathrm{H}_2)$, of $8\times 10^2$~cm$^{-3}$ and $4\times 10^2$~cm$^{-3}$, respectively (using $\rho = 1.4 m_\mathrm{H} n_\mathrm{H}$). In addition, for fast shocks the average pre-shock magnetic field strength $\left< B_0 \right> \sim 10 \, \mu$G, whereas for slow shocks $\left< B_0 \right> \sim 17 \, \mu$G. The average pre-shock density and magnetic field strength for fast shocks are within the ranges used by \cite{pon_molecular_2012} to model two-fluid C-type fast shocks. While the average density jump is much higher in slow shocks, $r\sim 15$, than in fast shocks, $r\sim 4$, the average post-shock densities are remarkably similiar: $\left< \rho_2 \right>_\mathrm{f,s} \sim 6\times 10^{-21}$~g/cm$^3$. This implies that both kinds of shocks are equally important with respect to star formation because it is the post-shock gas that sets the initial conditions for dense-core and star formation \citep{padoan_star_2011, federrath_star_2012, padoan_infall-driven_2014}.

These distributions suggest that to understand the impact that MHD shocks have on molecular clouds we need to understand both fast and slow MHD shocks. In a future paper we will apply this algorithm to other MHD simulations of molecular cloud turbulence \cite[e.g.,][]{federrath_star_2012, federrath_inefficient_2015}. We will investigate how the mixture of shock families may depend on the parameters of turbulence, e.g. the initial magnetic field strength, inclusion of extra physics such as self-gravity or protostellar jet feedback. If the mixture of shock types proves to be sensitive to these parameters, then differences in observational signatures between the shock types become signatures of these parameters. In Section~\ref{sec:heating} we discuss how one can use the shock mixture to compute the filling factor of hot, dense shocked gas.

\subsection{Energetics}

We may also consider the energetics of the shocks by comparing the kinetic energy dissipated in the shocks to the energy available in the turbulent motions. The kinetic flux through a unit area of shockfront is
\begin{align*}
K = \frac{1}{2}\rho_0 v_\mathrm{s}^3
\end{align*}
and the turbulent energy density is
\begin{align*}
\Gamma = \frac{1}{2}\left< \rho \right> \sigma^2
\end{align*}
where $\sigma$ is the velocity dispersion. Thus the timescale for dissipation in turbulence-driven shocks is
\begin{align*}
\tau_D = \Gamma V / \sum_K \left( K \left(\Delta x\right)^2 \right)
\end{align*}
where $V$ is the volume of the simulation ($8^3$~pc$^3$) and the summation is over the detected shocked cells. 

For the shocks shown in Fig.~\ref{fig:shockresults}, fast shocks dissipate $\sim 8$ times more energy than slow shocks due to the high velocity tail seen in Fig.~\ref{fig:histospeeds}. Note that not all of the kinetic energy from fast shocks dissipates by cooling, however, as some fraction goes into strengthening the magnetic field (as discussed in Section~\ref{sec:shocks}). Applying the MHD jump conditions for fast shocks with velocities ranging from 1--2~km/s, propagating into gas with the average fast pre-shock density $\left< \rho_0 \right>_\mathrm{f} \sim 2\times 10^{-21}$~g/cm$^3$ and average pre-shock magnetic field strength $\left< B_0 \right> \sim 10 \, \mu$G, we find that 35--70\% of the kinetic flux is lost to the magnetic field, depending on the orientation of the shock direction with respect to the magnetic field. As energy stored in the magnetic field is free to further dynamically impact the turbulence, we reduce the energy dissipated by fast shocks by a factor of 2 in order to capture the energy being dissipated as heat. 

Considering all shocks together, the shock dissipation rate is $\sim 0.2 \Lsun$. Compared to the turbulent kinetic energy, this gives a dissipation timescale of $\tau _D \sim 8$~Myrs. This is $\sim 1.3$ times the eddy turnover time $\tau_l=\left(\sqrt{3}L/2\right)/\sigma_\mathrm{1D}$ where $\sqrt{3}L$ is the size of the diagonal of the simulation and $\sigma_\mathrm{1D} = \sigma/\sqrt{3}$ is the one-dimensional velocity dispersion. From observations of turbulent dissipation regions, \cite{pon_mid-j_2014} estimated this ratio to be $1/3$ in the Perseus molecular cloud and \cite{larson_evidence_2015} made estimates of 0.94 and 0.65 for two shock models in the Taurus molecular cloud. Previous simulations have suggested that shocks dissipate around 50\% of the turbulent kinetic energy \citep{stone_dissipation_1998}. If we take this into account, our predicted ratio $\tau_D/\tau_l$ would be reduced by a factor of 2, placing it in the middle of these obversational results.

\section{Discussion}\label{sec:discussion}

We have presented an algorithm to detect and characterise fast and slow MHD shock waves in simulations of turbulent molecular clouds. While there is some observational evidence for the presence of fast \citep{lesaffre_low-velocity_2013, pon_mid-j_2014, larson_evidence_2015} and slow \citep{lehmann_signatures_2016} MHD shocks in molecular clouds, we present in this work the first prediction of the relative fraction of fast and slow shocks in molecular clouds. We characterised the shocks and provide the typical pre-shock conditions that should be used in future shock models that wish to model turbulence-driven MHD shocks. In the following we compare our work with other shock-finding algorithms, and then discuss how the results of the algorithm can be used to obtain an estimate of the volume of shock heated gas.

\subsection{Comparison to previous work}\label{sec:previous}
\cite{smith_shock_2000} and \cite{smith_distribution_2000} developed a method for counting shocks in MHD simulations of decaying and driven turbulence, respectively. Their method computes the velocity jump across converging regions. They found that the number distribution of these jumps did not substantially differ with the addition of a magnetic field. Our results are qualitatively similar, with weaker shocks dominating the number distribution, though they find much lower Mach numbers in general. This could be because they do not consider the shock reference frame, which introduces a correction to the velocity jump (see step~(v) of section~\ref{sec:alg_summary}). Their method does not explicitly disentangle the MHD shock types, and so cannot quantify the relative importance of fast or slow shocks.

The importance of shock heating on the chemical evolution of turbulent molecular clouds is highlighted by \cite{kumar_astrochemical_2013}. Like our work they capture the effects of shock heating on subgrid scales. They do this by post-processing Lagrangian tracer particles in a simulation of hydrodynamic turbulence. Their subgrid model is a one-dimensional integration of the fluid equations including a vast chemical network and molecular cooling. They were able to distinguish between chemicals that trace the mean physical state of cloud, and those that trace the non-equilibrium shock-heated gas. Their method also accounts for solenoidal heating and so they can measure the relative importance of these two heating mechanisms. However, as they only consider hydrodynamic turbulence, their results are not sensitive to the distinct effects of MHD shock types. Extending their work to the MHD case is in principle simple. The subgrid model would need to include MHD effects like the shock models of \cite{flower_excitation_2010}, \cite{pon_molecular_2012} or \cite{lehmann_signatures_2016}. Some difficulty lies in obtaining the pre-shock state, which requires knowledge of the magnetic field direction with respect to the shock propagation direction. In addition, the pre-shock state does not uniquely determine the MHD shock type \citep[cf][]{kennel_mhd_1989}. We have addressed this problem by using both pre- and post-shock information in order to ascertain two of the three possible shock types. 

The shock finding algorithm most similar to \textsc{shockfind} is that outlined in \cite{schaal_shock_2015}. They look for shocks in cosmological hydrodynamic simulations. Their method flags cells of converging flow, and defines the shock direction using the gradient of the temperature. They then use a series of criteria to filter spurious shock detections. While their work does not include magnetic fields, and thus does not consider different shock families, it would be simple to extend their algorithm to do so. It would only take the addition of further filtering criteria such as we presented in step~(vi) of section~\ref{sec:alg_summary}. This extension would allow for a comparison of our work to MHD simulations using moving-mesh codes such as \textsc{arepo} \citep{springel_e_2010, pakmor_magnetohydrodynamics_2011}.

\subsection{Shock heating}\label{sec:heating}

\cite{lehmann_signatures_2016} showed that C-type fast MHD shocks and J-type slow MHD shocks distinctly heat the gas they propagate through. The fast shocks modeled in \citeauthor{lehmann_signatures_2016} reach peak temperatures of $\sim 150$~K, whereas slow shocks could reach temperatures of $\sim 800$~K. This is because in the weakly ionized gas that makes up molecular clouds, the heating timescale in fast shocks is determined by the ion-neutral collision timescale, which is slower than the cooling timescale. In contrast, in slow shocks, the ion-neutral collision timescale is shorter than the cooling timescale, such that heating in slow shocks is more significant. Even though these high temperatures only occupy a thin shock layer, the heating is important because of the rich chemistry it activates. The chemical signatures of shocks may persist in regions of unshocked gas. In this section we use the pre-shock conditions (Fig.~\ref{fig:BversusRHO}) and shock front area (Fig.~\ref{fig:histospeeds}) obtained by \textsc{shockfind}, combined with representative sub-grid two-fluid shock models from \cite{lehmann_signatures_2016} to estimate the volume of shocked gas in a turbulent cloud.

\subsubsection{Sub-grid two-fluid shock models}

Ion-neutral collisions determine shock thickness in two-fluid shocks, and so the ionization fraction is a key variable in determining the volume filling fraction of shocks. Ionization sources, such as cosmic-rays and ultraviolet photons, are density dependent and so the ionization fraction spatially varies in a turbulent cloud. For simplicity we adopt the ionization fraction of \cite{bergin_cold_2007}: $x_e = 1.3 \times 10^{-5} n(\mathrm{H}_2)^{-1/2}$ where we use the density in the pre-shock gas. This leads to pre-shock ionization fractions ranging between $2\times 10^{-7}$ and $9\times 10^{-6}$.

\begin{figure}
  \centering
    \includegraphics[width=0.5\textwidth]{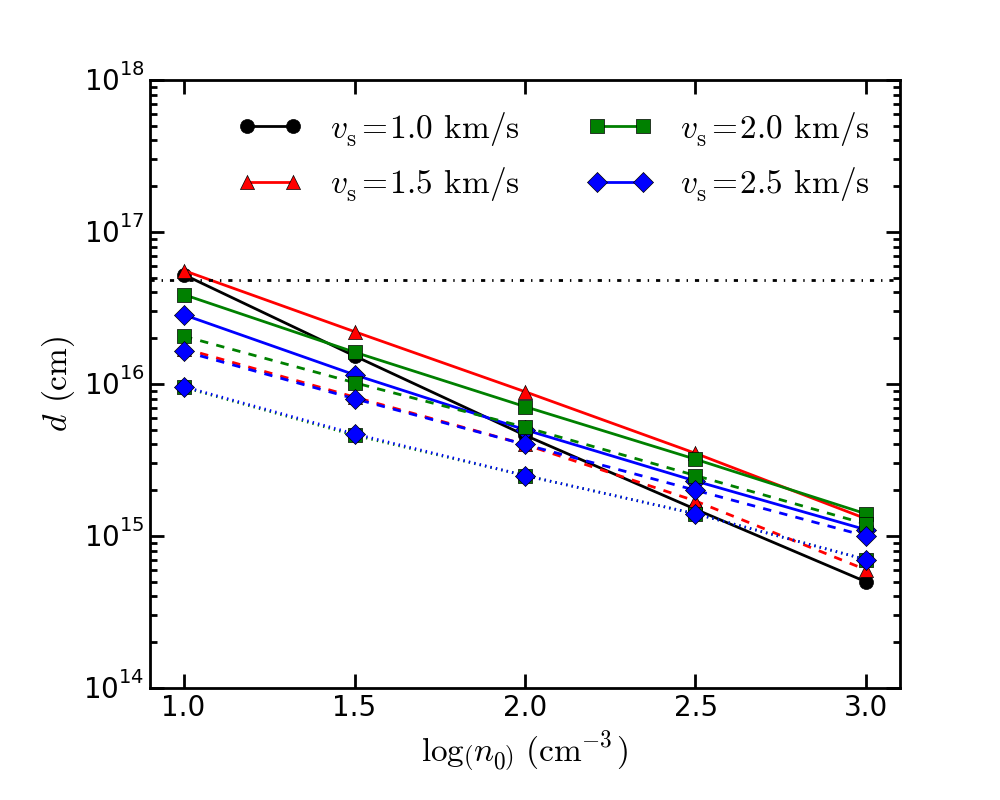}
    \caption{Thickness of slow MHD shocks versus preshock total hydrogen density. The solid, dashed and dotted lines show shock thicknesses above 50, 100 and 150 K, respectively. Each line of the same colour (or marker) represents models with the same shock velocity. The horizontal dash-dotted line shows the size of a cell in the turbulent cloud simulation.}
  \label{fig:thickness}
\end{figure}

For slow MHD shocks, the shock thickness is independent of the pre-shock magnetic field strength. So, for a given pre-shock density, we choose the magnetic field such that the Alfv\'{e}n velocity $v_\mathrm{A} = 3$~km/s. This allows us to compute slow shocks with speeds up to $v_\mathrm{A} \cos \theta$, where $\theta$ is the angle between the direction of propagation and pre-shock magnetic field. The dashed line in Fig.~\ref{fig:BversusRHO} traces the pre-shock magnetic field and mass density at this Alfv\'{e}n velocity, which roughly follows the distribution of slow shocks in the simulation. We model slow MHD shocks for $\theta=30^\circ$ allowing shock velocities up to 2.5~km/s. The number of slow shocks above this velocity in the simulation is only $\sim 0.7$\% of all slow shocks (see Fig.~\ref{fig:histospeeds}), so our results are only negligibly affected by this limit. Fig.~\ref{fig:thickness} shows the thicknesses of slow MHD shocks heated above 50~K and 100~K, for pre-shock total hydrogen densities ranging between 10 and $10^3$~cm$^{-3}$. The hot shock front is largest for models with the lowest pre-shock density and lowest velocity, peaking at $\sim 6\times 10^{16}$~cm which is of the order of the size of a cell $\Delta x$. This is important because it means that the substructure within the shock front would not dynamically affect the scales that the simulation captures.

For fast MHD shocks with a given pre-shock density, we choose the magnetic field such that the Alfv\'{e}n velocity $v_\mathrm{A} = 1$~km/s. The solid line in Fig.~\ref{fig:BversusRHO} traces this Alfv\'{e}n velocity, which roughly follows the distribution of fast shocks in the simulation. The shock thickness, $d$, of two-fluid fast C-type shocks is estimated as
\begin{align*}
d \sim \frac{v_\mathrm{s}}{n_i \alpha}
\end{align*}
where $n_i = x_\mathrm{e} n_\mathrm{H}$ is the number density of ions and $\alpha = 1.6 \times 10^{-9}$~cm$^3$/s is the rate coefficient for ion-neutral scattering. This shock thickness estimate can exceed the simulation box size at low densities and large shock velocities, which means, of course, that some fast shock models are innapropriate as sub-grid models in this ideal MHD simulation. Thus we consider models with thickness $d \leq 10\Delta x$ as small enough to not significantly affect the simulation results. Note that this estimate gives the thickness of steady-state fast shocks. Factoring in the structure of non-steady shocks is beyond the scope of this work. Finally, we find that models of fast shocks for these Mach numbers and densities in the range $n_\mathrm{H}=10$--$10^3$~cm$^{-3}$ are all C-type shocks.

\begin{figure*}
  \centering
    \includegraphics[width=\textwidth]{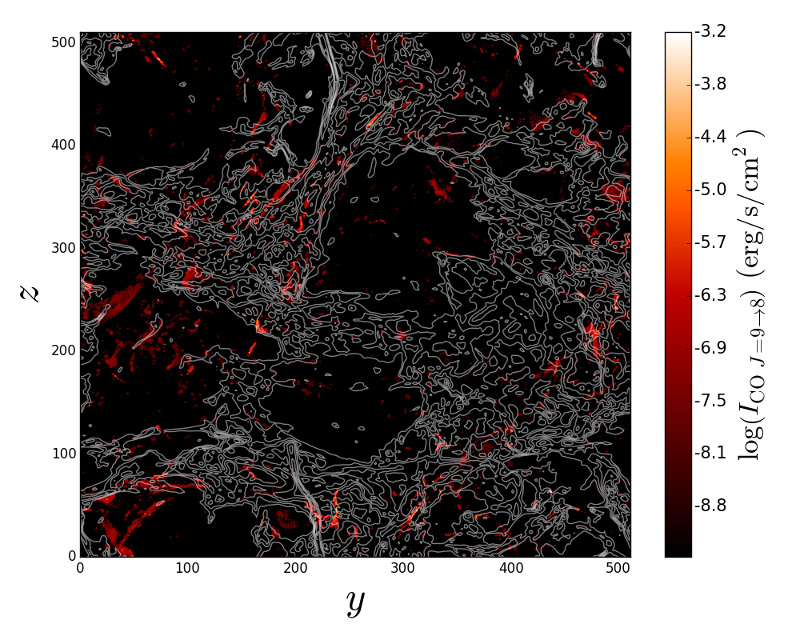}
    \caption{Predicted synthetic radio emission map of the simulation in the CO $J$=9--8 rotational transition, computed with \textsc{radex}. The white line contours are of the total hydrogen column density, equally spaced from the average column over the whole face ($\left< N_\mathrm{H} \right> \sim 10^{24.5}$~cm$^{-2}$) up to the maximum column ($N_\mathrm{H,max} \sim 10^{25.5}$~cm$^{-2}$).}
  \label{fig:COmap}
\end{figure*}

We bin the detected shocks into pre-shock total hydrogen densities centred on $10^1$, $10^{1.5}$, $10^{2.0}$, $10^{2.5}$ and $10^{3.0}$~cm$^{-3}$. These values of density almost cover the entire range of detected shocks, with $\lesssim 5\%$ of slow shocks and $\lesssim 2\%$ of fast shocks falling outside. With this binning, we can use the shock thicknesses from Fig.~\ref{fig:thickness} to estimate the volume of warm shocked gas. Using the area computed for Fig.~\ref{fig:histospeeds} multiplied by the shock thicknesses we find that $f_\mathrm{T>50\,K} \sim 0.03\%$ of the volume is filled with shocked gas greater than 50~K. This is of the order of the shocked volume filling factor measurement of \cite{pon_mid-j_2014} from turbulent dissipation regions in the Perseus molecular cloud. In addition, $f_\mathrm{T>100\,K} \sim 5 \times 10^{-3}\%$ and $f_\mathrm{T>150\,K} \sim 9\times 10^{-4}\%$ of the volume is filled with shocked gas greater than 100~K and 150~K, respectively. This warm gas occurs entirely in slow shocks, because the fast shocks at these conditions do not reach peak temperatures above 50~K. Hence, if no distinction of MHD shock families is made and all shocks in an MHD simulation were assumed to be fast C-type shocks, there would be no warm component of gas with temperature $T>50$~K at all.

\subsubsection{Rotational line emission}

While this predicted filling factor of gas hotter than 50~K, $\sim 0.03\%$, is very small, only $\sim 2.5\%$ of the volume is filled with gas at densities higher than the average post-shock density $\left< \rho_2 \right> \sim 6\times 10^{-21}$~g/cm$^3$. As an example observational impact of this warm gas, we estimate the intensity of a CO rotational line using the non local thermodynamic equilibrium (LTE) radiative transfer code \textsc{radex} \citep{van_der_tak_computer_2007}. 

For a given radiating molecule, \textsc{radex} requires as input the density of H$_2$ as the collisional partner, the column density of the radiating molecule, the temperature and the linewidth. In order to avoid geometrical effects, we consider only optically thin lines. By doing this, we can use \textsc{radex} in slab mode and treat each column through the simulation as consisting of a simple addition of slabs. We use the shock thicknesses computed above to define at each cell (on one face) the column density excited by gas at 75, 125 and 175 K. We use a CO abundance of $x(\mathrm{CO}) = n(\mathrm{CO})/n_\mathrm{H} = 1.2\times 10^{-4}$ to derive the CO column density. We then take the density of H$_2$ to be equal to the average post-shock density. We assume that the rest of the gas makes up a column of CO excited by 10 K gas at an H$_2$ density equal to the average density in the simulation. Finally, we use the velocity dispersion of the cloud as the input linewidth.

Fig.~\ref{fig:COmap} shows the synthetic map of \textsc{radex} estimated CO $J$=9-8 intensities with contours of total hydrogen column density overlaid. We chose the $J$=9-8 line because it was the lowest $J$ CO line that was optically thin at the column densities reached here and because another source of high-$J$ CO lines, photodissociation regions, may have difficulties producing significant emission at this line and above \citep{pon_molecular_2012}. The emission in Fig.~\ref{fig:COmap} is entirely due to shocks---as the background 10 K gas negligibly emits in this line---and is strongest in filamentary structures. This is because an edge-on shock, with respect to the line of sight, presents a larger column of heated gas than a face-on shock. These filamentary emission regions also tend to occur in regions with large hydrogen column densities, suggesting that the pre-shocked gas is already at a high density. The correlation between large hydrogen column density and emission is not perfect, however, as there is some significant CO emission at regions of below average hydrogen column density. If we add up the emission from over the whole face of the cloud, then the total cloud luminosity at this line is $\sim 4\times10^{-3}\Lsun$. Notably, if we ignored the distinction of MHD shock families and assumed that all shocks were the well-studied C-type fast shocks, we would predict that the CO $J$=9-8 emission would be negligible.

Estimates of high-$J$ CO lines like this could provide distinct observational predictions between different simulations of turbulent clouds. The accuracy of this estimate depends on the accuracy of the estimate of the volume of warm gas. This was estimated using the shock thicknesses derived from the two-fluid shock models of \cite{lehmann_signatures_2016}. It also only included the gas heated by slow shocks, because some two-fluid C-type fast shocks have thicknesses too large to be applicable to this simulation and the remaining fast shocks do not reach peak temperature of 50~K. This implies that a large proportion of the fast shocks detected here would not have the steady-state structure of two-fluid fast shocks. We have also ignored the possibility of intermediate MHD shocks, because they do not have the predictable impact on magnetic fields that this algorithm exploits. In addition, the shock models of \cite{lehmann_signatures_2016} are highly simplified in order to highlight the differences between fast and slow shocks. Improvements in models of shocks, such as using an expanded chemical network and including the effects of dust grains, would improve the accuracy of the shock-heated volume estimate. 

\section{Conclusion}\label{sec:conclusion}

\highlight{The publicly available algorithm \textsc{shockfind}\footnote{Found on BitBucket (https://bitbucket.org/shockfind/shockfind) and the \textsc{python} Package Index (https://pypi.python.org/pypi/shockfind)} was developed, which} extracts and characterises the shock waves in MHD simulations. This algorithm was applied to a high-resolution simulation of a magnetised, turbulent molecular cloud. We presented the first prediction of the relative fraction of fast and slow MHD shocks in this turbulent molecular cloud. The sonic and Alfv\'{e}nic Mach number distributions for these two families of shocks are distinct and confirm that low-velocities, below $v_\mathrm{s} = 3$ km/s, dominate the population of shocks. By considering the energetics of the detected shocks, we found that the ratio of the shock dissipation timescale to cloud crossing time is comparable to observed values from turbulent dissipation regions in molecular clouds. We have also used simple sub-grid models of two-fluid MHD shocks from \cite{lehmann_signatures_2016} to estimate the heating that would occur within the thin shock front of these shocks. Slow MHD shocks were found to produce a low volume filling factor, $\sim 0.03 \%$, component of the cloud heated above 50 K with a small portion of this component reaching temperatures above 150 K. 

We used the non-LTE radiative transfer code \textsc{radex} to estimate the intensity of a high-$J$ CO rotational transition and found that the shock-heated gas radiates far above the background cloud intensity. High-$J$ CO line emission may therefore be an important observational diagnostic of shocks in molecular clouds.

Our shock-detection algorithm is general enough to be applied widely to MHD simulations of other astrophysical phenomena. It would be interesting to see the mixture of shock families that might be present in simulations of supernova shocks, protostellar jets interacting with the interstellar medium, colliding flows, cloud-cloud collisions, etc. In a future paper we plan to extract and characterise the MHD shocks in a variety of simulations of turbulent molecular clouds in order to search for correlations between the parameters of turbulence and possible observational effects of MHD shock waves.

\section*{Acknowledgements}
The authors gratefully acknowledge discussions with James Tocknell and Birendra Pandey. This research made use of Astropy, a community-developed core Python package for Astronomy \citep{2013A&A...558A..33A}. A.L.~was supported by an Australian Postgraduate Award. 
C.F.~acknowledges funding provided by the Australian Research Council's Discovery Projects funding scheme (grants~DP130102078 and~DP150104329) and M.W.~acknowledges funding provided by the Australian Research Council's Discovery Projects funding scheme (grant~DP120101792).
We gratefully acknowledge the J\"ulich Supercomputing Centre (grant hhd20), the Leibniz Rechenzentrum and the Gauss Centre for Supercomputing (grants~pr32lo, pr48pi and GCS Large-scale project~10391), the Partnership for Advanced Computing in Europe (PRACE grant pr89mu), the Australian National Computational Infrastructure (grant~ek9), and the Pawsey Supercomputing Centre with funding from the Australian Government and the Government of Western Australia.
The software used in this work was in part developed by the DOE-supported Flash Center for Computational Science at the University of Chicago.

\bibliographystyle{mn2e}
\bibliography{shockextraction}

\appendix
\section{Cylinder Radius} \label{ap:cylinder}

\begin{figure}
  \centering
    \includegraphics[width=0.5\textwidth]{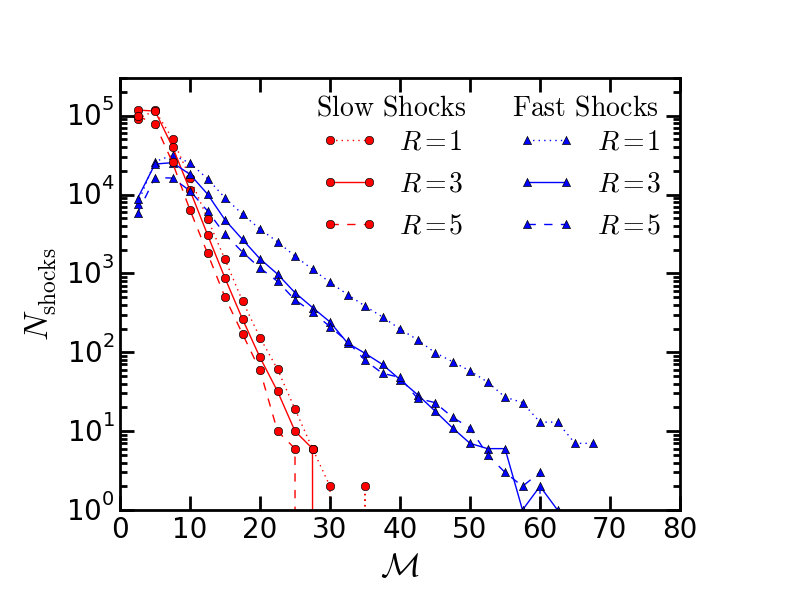}
    \caption{Mach number distributions for the search described in Section~\ref{sec:turbulence} but with averaging cylinder radiis of $R=1$ (dotted), 3 (solid) and 5 (dashed).}
  \label{fig:cylinder}
\end{figure}

Here we check the effect of varying the radius of the cylinder used to define average pre- and post-shock variables (step (iii) in Section~\ref{sec:alg_summary}). Fig.~\ref{ap:cylinder} shows the Mach number distributions for fast and slow shocks for runs of \textsc{shockfind} on the same locations, but with cylinder radii of $R=1$, 3 and 5 in units of cell size. A radius of $R=1$ defines the minimum cylinder size to represent a line that does not suffer aliasing effects. This size, however, is still affected by small scale numerical noise. Averaging over larger radii, $R=3$ and $R=5$, removes the effect of the numerical noise. There is little difference between these two larger radii, and so we choose $R=3$ as the compromise between noise and lowering computational time.

\section{Overcounting the shocked cells} \label{ap:overcounting}

\begin{figure}
  \centering
    \includegraphics[width=0.5\textwidth]{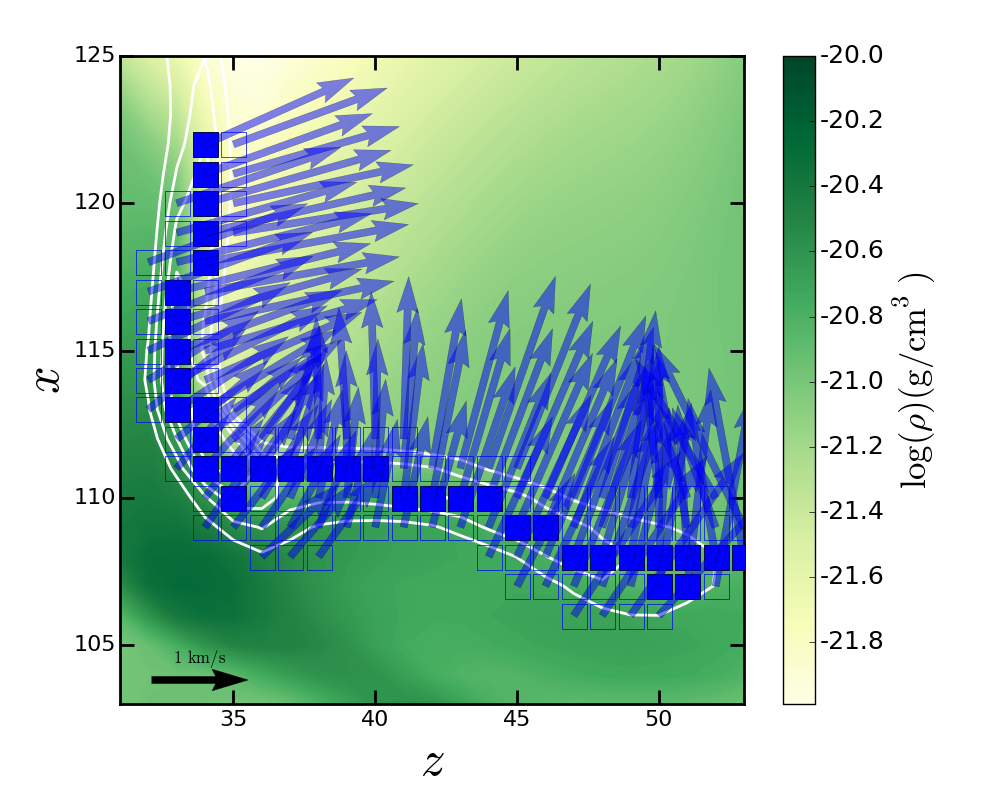}
    \caption{Constant y slice of mass density (filled contours), convergence (white contours), detected shocked cells (squares) and arrows proportional to shock velocity in this plane. The filled squares denote cells at convergence peaks along the line defined by the shock velocity vector at that cell.}
  \label{fig:double}
\end{figure}

In Fig.~\ref{fig:double} we show a close up of the detected shocked cells of a curved shock front. In this figure, it can be seen that the thickness of the shock front is 3 or 4 rows of detected cells. As the shock front area is required to link our work to the results of one-dimensional shock models, we are overcounting if we consider every detected cell as a unique unit of shocked area. In order to avoid this over counting we only use (as step (viii) of Section~\ref{sec:alg_summary}) cells that occur at the local convergence maxima (filled squares in Fig.~\ref{fig:double}) along the extracted line (step (iii) of Section~\ref{sec:alg_summary}).

\section{Convergence Test} \label{ap:convergence}

\begin{figure}
  \centering
    \includegraphics[width=0.5\textwidth]{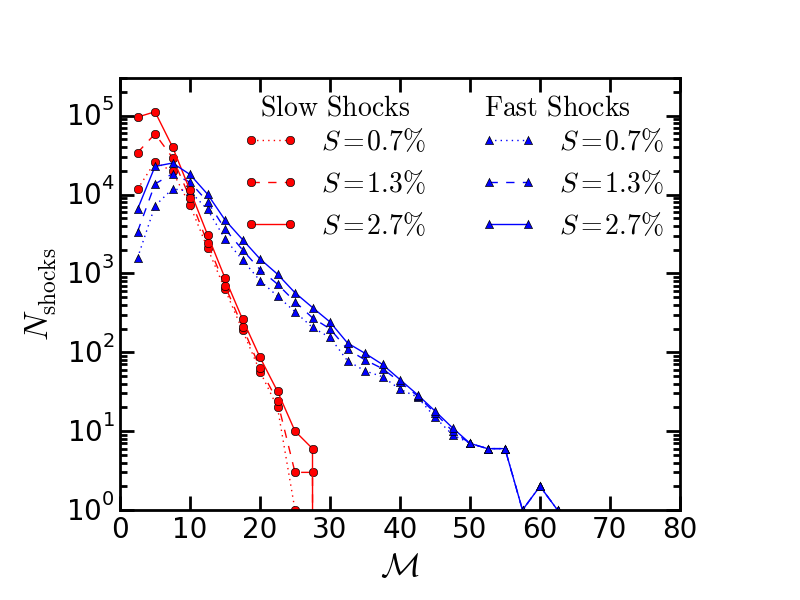}
    \caption{Mach number distributions for searches of $S=0.7\%$, $1.3\%$, and $2.7\%$ of possible shocked cells (those with $\nabla \cdot \mathbf{v} < 0$).}
  \label{fig:resolution}
\end{figure}

Here we check that the search thresholds capture converged distributions of shocks. The convergence threshold defined in Sec.~\ref{sec:thresholds} searches down to $S=2.7\%$ of cells with negative divergence ($\nabla \cdot \mathbf{v}$), i.e., of all possible shocked cells. Fig.~\ref{fig:resolution} shows the effect of doubling the number of cells searched on the Mach number distributions of fast and slow shocks. \highlight{The distributions are converged for the most observationally important shocks ($\mathcal{M} > 5$)}.

\end{document}